\documentclass[preprints,article,accept,moreauthors,pdftex]{Definitions/mdpi}

\firstpage{1} 
\pubvolume{11}
\issuenum{29}
\articlenumber{1}
\pubyear{2020}
\copyrightyear{2020}
\history{\small{\today \, UTC \currenttime}}
\usepackage{gensymb}
\usepackage{amssymb}
\usepackage{amsfonts}
\usepackage{miller}
\usepackage{nicefrac}
\usepackage{siunitx}
\usepackage{comment}
\usepackage{lineno}
\usepackage[warn]{textcomp}
\usepackage[section]{placeins}
\usepackage{upgreek}
\usepackage{animate}
\usepackage[yyyymmdd,hhmmss]{datetime}

\usepackage{url}
\usepackage{geometry}
\Title{Fixing Left and Right: Assignment of Chiral Elemental Crystal Structures using Kikuchi Diffraction}

\Author{Aimo Winkelmann $^{1,2}$\orcidA{}, Grzegorz Cios $^{1}$\orcidC{},
Tomasz Tokarski $^{1}$\orcidT{}, \mbox{Piotr Bała $^{1,3}$\orcidP{}}, Yuri Grin$^{4}$, Ulrich Burkhardt$^{4}$}

\AuthorNames{Aimo Winkelmann,  Grzegorz Cios,
Tomasz Tokarski, Piotr Bała, Yuri Grin, Ulrich Burkhardt}

\address{%
$\texorpdfstring{^{1}}{Lg}$ \quad Academic Centre for Materials and Nanotechnology, AGH University of Science and Technology, al. A. Mickiewicza 30, 30-059 Krak\'{o}w, Poland\\
$\texorpdfstring{^{2}}{Lg}$ \quad Department of Physics, SUPA, University of Strathclyde, Glasgow G4 0NG, United Kingdom\\
$\texorpdfstring{^{3}}{Lg}$ \quad Faculty of Metals and Industrial Computer Science, AGH University of Science and Technology, al. A. Mickiewicza 30, 30-059 Krak\'{o}w, Poland\\
$\texorpdfstring{^{4}}{Lg}$ \quad 
Max-Planck-Institut für Chemische Physik fester Stoffe, 01187 Dresden, Germany\\}

\corres{Correspondence: winkelmann@agh.edu.pl (A.W.)}

\abstract{
Crystals of the chemical elements manganese, tellurium, and selenium can show the effects of handedness.
In order to sense the possible effects of a changing sense of chirality on the properties of samples from these elements, the potential presence of two, enantiomorphic, physically different, variants of these elemental crystal structures needs to be resolved in crystallographic analyses.
Due to fundamental limitations of kinematical X-ray scattering in crystals, however, the effects of chirality in single-element crystals are very difficult to sense using standard X-ray diffraction techniques.
In the present paper, we show that dynamical Kikuchi diffraction in the scanning electron microscope is sensitive to the local sense of chirality in crystals of single chemical elements.
We demonstrate chirality assignment in $\beta$-manganese, and we determine the sense of crystal chirality from Kikuchi diffraction patterns of the trigonal structures of tellurium and selenium.
}

\keyword{chirality; non-centrosymmetric crystals; scanning electron microscopy; electron backscatter diffraction; Kikuchi diffraction; absolute structure}

\begin{document}

\section{Introduction}

\begin{figure}[!htb]
  %\centering
  \hspace{-1.8cm}
  \includegraphics[width=18cm]{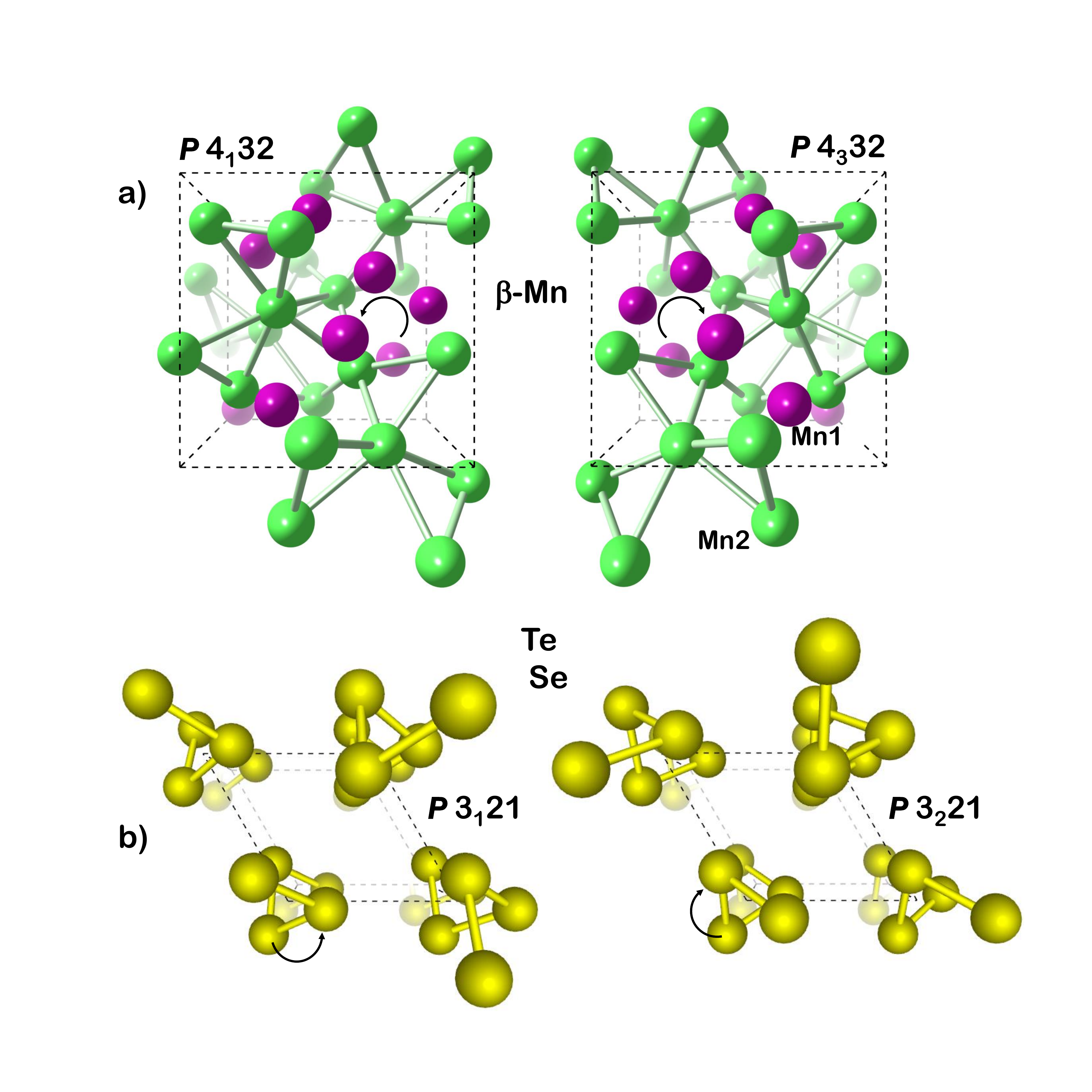}
   \caption{a) Structure of the two enantiomorphs of $\beta$-Mn. The $\beta$-Mn crystal structure contains two different types of sites: Mn1 (purple) and Mn2 (green). The "twisted windmill"  arrangement \cite{hafner2003prb2} of the Mn2 atoms along a left-handed or right-handed screw axis is used to emphasize the mirror relationship between the two enantiomorphs of space groups $P\,4_132$ and $P\,4_332$.
   b) The chiral crystal structures of trigonal Te and Se are characterized by spirals of different handedness along the $c$-axis of the unit cell in the space groups $P\,3_121$ and $P\,3_221$.}
\label{fig:mnunit}
\end{figure}

Throughout the periodic table, most of the pure chemical elements crystallize into simple structures which contain a center of symmetry, relative to which an inversion of all atomic positions $(x,y,z)$ by $(-x,-y,-z)$ has no effect. 
The elements of manganese, selenium, and tellurium, however, are exceptions to this rule because they can form chiral crystal structures which do not contain any mirror planes or inversion centers \cite{flack2003hca}. 
The inversion of the atomic coordinates in such crystals will lead to a physically different, enantiomorphic, structure which looks like the mirror image of the initial one.
This is illustrated in Figure\,\ref{fig:mnunit}(a) for the cubic crystal structure of $\beta$-Mn \cite{shoemaker1978acb}, and in Figure\,\ref{fig:mnunit}(b) for the trigonal structures of Te and Se \cite{cherin1967aca,cherin1967ioc}.
In principle, any polycrystalline sample of these chiral element structures can thus potentially contain two enantiomorphic variants, which can affect the physical and chemical properties of the material \cite{newnham2005properties,barron2009chapter}.

The unit cell of $\beta$-Mn  exhibits a complex arrangement of 20 atoms, and the two enantiomorphic structures belong to the two different space groups $P\,4_132$ and $P\,4_332$ \cite{shoemaker1978acb}. 
These can be discriminated in Figure\,\ref{fig:mnunit}(a) by an opposite handedness of the four-fold screw axes ($4_1$ vs. $4_3$). 
The fundamental $\beta$-Mn structure type, and its possible realization in two possible space groups, is well known from X-ray diffraction (XRD) studies \cite{preston1928pma,shoemaker1978acb}. 
The physics of kinematical X-ray scattering, however, severely limits the actual precise assignment of the sense of chirality of the specific, mirror-related, variants in elemental crystals \cite{glusker2010}.  
In consequence, single-crystal, unpolarized X-ray diffraction patterns which are conventionally used for absolute structure determination  \cite{flack2008chirality} will remain essentially indistinguishable for enantiomorphs of chiral structures which contain only scattering atoms of the same type, even when anomalous dispersion effects are taken into account \cite{tanaka2010jpcm}. 
Basically, kinematical unpolarized X-ray scattering allows to conclude that only two different modifications of a chiral elemental structure are compatible with the measured diffraction intensities, but it cannot reveal how much of each possible variant is actually present in a given sample. 
As an additional problem for $\beta$-Mn, large single crystals are not available, because $\beta$-Mn is a high-temperature phase which is thermodynamically stable between 727\,$^\circ$C and 1095\,$^\circ$C, while being metastable at room temperature \cite{shoemaker1978acb}.
Because of these various complications, an adequate microcrystallographic discrimination method between the two different enantiomorphs of $\beta$-Mn in real materials was demonstrated only recently in \cite{burkhardt2021sciadv}, almost 100 years after the first successful structure determination of $\beta$-Mn \cite{preston1928pma}. 
In \cite{burkhardt2021sciadv}, it was shown how dynamical electron diffraction in SEM Kikuchi patterns can overcome key limitations of conventional X-ray diffraction, allowing to obtain precisely defined microstructural chirality information from polycrystalline samples of a key chemical element like manganese.

The study of the breaking of inversion symmetry in $\beta$-Mn is particularly important because $\beta$-Mn is also a prototype structure for a range of binary and ternary compounds with complex magnetic effects \cite{nakamura1997pb,hafner2003prb2,eriksson2005prb,paddison2013prl,tokunaga2015ncomm}.
This includes the formation of skyrmions which show an internal sense of rotation that is linked to the sense of chirality of the host lattice \cite{roesler2006nat,nagaosa2013nnano,karube2018prb}. 
Spin-dependent topological effects are discussed to occur in the electronic structure of chiral crystals with a $\beta$-Mn-derived structure \cite{chang2018nmat}.
Chirality-resolved studies could thus allow to separately address the breaking of inversion symmetry in pure $\beta$-Mn-type crystals, independently of any additional effects which are due to the different chemical occupation of the atomic sites in the crystal unit cells of isotypic binary or ternary phases.

Closely related problems of an efficient assignment of the sense of chirality as for $\beta$-Mn do also exist for the chiral element structures of tellurium and selenium. 
These elemental modifications can be observed in a trigonal crystal structure \cite{cherin1967aca,cherin1967ioc} with the enantiomorphs belonging to space group $P\,3_121$ and $P\,3_221$, respectively. 
Due to its high atomic number, notably tellurium is discussed to exhibit chirality dependent effects of spin-orbit coupling in its electronic structure \cite{tsirkin2018prb,ideue2019pnas,sakano2020prl}. 
Based on the interplay of a chiral crystal structure and the polarization state of the scattered radiation \cite{tanaka2008prl}, chirality determination of macroscopic Te single-crystals has been achieved using circularly polarized X-rays in synchrotron-based diffraction experiments \cite{tanaka2010jpcm}, as well as by polarized neutron diffraction \cite{brown1996aca}. 
The use of polarized X-ray diffraction for discrimination of the effects of chirality in $\beta$-Mn-type crystals has been analyzed theoretically in \cite{lovesey2021prb}.

Circumventing some of the limits of X-ray diffraction, electron diffraction methods can grant access to the chirality of crystal structures based on the much stronger interaction of electrons with matter. %\cite{fultz2008tem}.
In this case, dynamical, multiple scattering, electron diffraction effects provide sensitivity to the sense of chirality in crystals \cite{goodmansecomb1977aca,spence1994aca}.
This has been demonstrated experimentally for a number of chiral crystal structures using different techniques in transmission electron microscopy (TEM) \cite{tanaka1985jpsj,ma2017nmat,dong2020ncomm,brazda2019science}, including atomic resolution of handedness in tellurium \cite{dong2020ncomm}.

In the scanning electron microscope (SEM), for comparison, Kikuchi diffraction is a particularly attractive solution because of its widespread availability in the commercial technique of electron backscatter diffraction (EBSD) \cite{schwartz2009ebsd2}, often with less demanding experimental requirements compared to TEM. 
Kikuchi diffraction patterns are formed from divergent sources of inelastically scattered  electrons, which are subsequently diffracted by the surrounding the crystalline volume \cite{schwartz2009ebsd2}, and it is well known that Kikuchi diffraction patterns are sensitive to the absence of a center of symmetry in crystal structures based on dynamical electron diffraction effects \cite{babakishi1989sc,winkelmann2015apl}.
The application of EBSD in the SEM is especially useful for microstructural analysis applications because it provides precise crystallographic measurements of extended polycrystalline sample areas up to the order of several square millimeters at simultaneously possible sub-micron spatial resolution,
with Transmission Kikuchi Diffraction (TKD) even providing spatial resolutions down to the order of 10\,nm \cite{sneddon2016mse}.
Quantitative Kikuchi diffraction analysis of materials can be based on the numerical comparison of experimental data with theoretical pattern simulations \cite{winkelmann2016iop}.
With respect to non-centrosymmetric crystal structures, quantitative Kikuchi diffraction based on dynamical diffraction simulations has provided access to the absolute polarity of III-V semiconductor compounds, which was confirmed by a comparison to the orientation of microscopic etch-pit features on the sample surface \cite{nolze2015jac}.  
Kikuchi patterns were also shown to be sensitive to the handedness of $\alpha$-quartz, which can be determined from the external shape or optical rotation measured for the respective macroscopic single crystals \cite{winkelmann2015um}.  
Ferroelectric domains in non-centrosymmetric LiNbO$_3$ have been distinguished based on a corresponding symmetry breaking of Kikuchi pattern features \cite{burch2017um}.
Finally, in combination with anomalous scattering in X-ray diffraction, Kikuchi diffraction from non-centrosymmetric crystals can be consistently calibrated for absolute structure determination of chiral intermetallic compounds \cite{burkhardt2020srep}.
These different studies showed that Kikuchi diffraction is an expedient method for the microscopic determination of absolute polarity and chirality in the SEM, which often can involve analytic tasks where other crystallographic analysis methods, such as XRD or TEM, are limited, or are difficult to apply.

In the present paper, we apply Kikuchi diffraction in the scanning electron microscope to discriminate between enantiomorphs of $\beta$-manganese, tellurium and selenium. 
We discuss in detail the chirality-dependent features in individual Kikuchi patterns of the two different enantiomorphs of $\beta$-Mn, and we demonstrate the assignment of the sense of chirality from Kikuchi diffraction patterns of Te and Se by a quantitative pattern matching approach.
These results provide the general basis for EBSD-based local chirality analyis in polycrystalline materials of Mn, Te, and Se.

\section{Methods}
\label{sec:methods}

\subsection{Sample Preparation}

For the Kikuchi diffraction pattern analysis discussed in Section \ref{sec:patterns111}, a piece of electrolytically purified manganese sized about 3mm $\times$ 10mm $\times$ 10 mm was placed in an air atmosphere furnace at 1000\,$^\circ$C and heated for 1 hour, after which it was quenched in water.
For the EBSD investigation, a standard mechanical preparation was used, grinding and polishing with diamond suspensions and colloidal silica for 20 minutes at the end.

Tellurium crystals were prepared in a closed quartz ampule by a chemical transport reaction with a temperature gradient between 375\,$^\circ$C (source) and 345\,$^\circ$C (sink). 
Polycrystalline material (Alfa Aesar 99.999\%) was used as source. 
The transport agent iodine in a concentration of 1.5\,$mg/cm^3$ (Alfa Aesar 99.998\%) supported the formation of cuboidal shaped crystals \cite{schaefer1991zaac}. In a subsequent EDX analysis of the grown Te crystals, iodine was below the detection limit.

Crystals of selenium were grown via vacuum-sublimation in a temperature gradient from 275\,$^\circ$C to 150\,$^\circ$C starting from liquid selenium (Alfa Aesar 99.999+\%).
The crystallization experiment was carried out in a horizontally arranged two-zone furnace which was tilted by approx. 20 degrees.

\subsection{Electron Backscatter Diffraction}

\subsubsection{Data Acquisition}

The experimental backscattered Kikuchi diffraction patterns
were acquired using a Quantax EBSD system (Bruker Nano, Germany) attached to a scanning electron microscope JEOL JSM7800 F, as well as by a Hikari EBSD system (EDAX Inc., USA) on a Versa 3D FE-SEM (FEI).
All measured Kikuchi patterns were saved for post-processing using custom data analysis software. 

\subsubsection{Pattern Matching Approach}

A best-fit pattern matching approach was used to find the orientation and the chirality corresponding to each experimental Kikuchi pattern \cite{nolze2018amat,winkelmann2020jm,winkelmann2020mat}. 
The simulated Kikuchi patterns were calculated using the Bloch wave approach \cite{winkelmann2009ebsd2}.
The normalized cross-correlation coefficient (NCC) \cite{metcalfe1994book}  $r$ ($0 \leq |r| \leq 1$) between the experimental and the simulated Kikuchi pattern of the reference structure was optimized, with an additional test against the inverted structure to discriminate between the enantiomorphs as described in detail in \cite{burkhardt2020srep}.

\section{Results and Discussion}

\subsection{Chirality dependent Kikuchi diffraction pattern matching} 
\label{sec:patterns111}

\begin{figure}[!tbhp]
	%\centering
	\hspace{-1.5cm}
	\includegraphics[width=18cm]{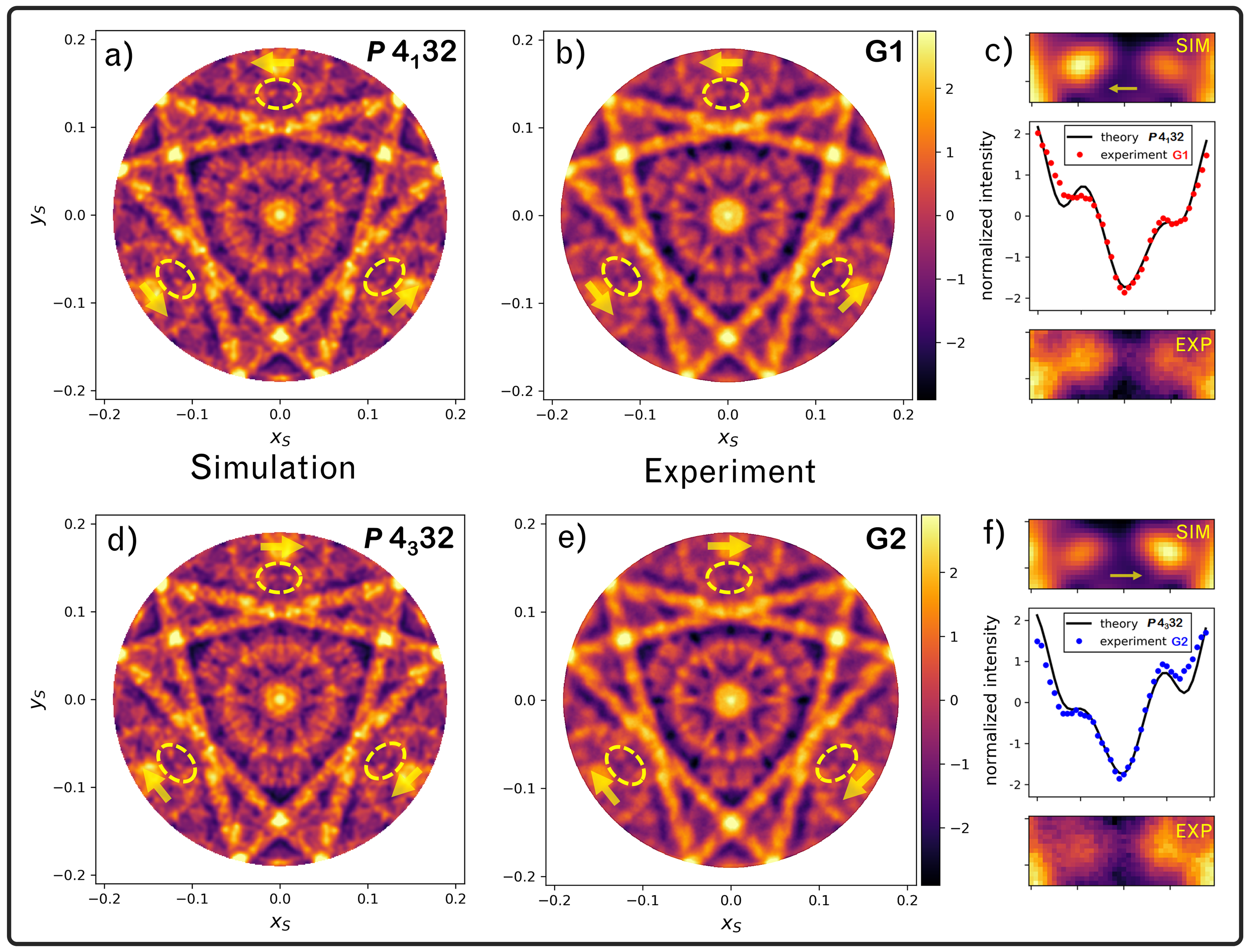}
	\caption{Assignment of the $\beta$-Mn enantiomorphs by comparison of simulated (a,d) to experimental (b,e) Kikuchi diffraction patterns (stereographic projection, electron energy 20\,keV).
	As shown by the arrows, the asymmetric features in the dashed ovals define the experimental sense of rotation, clockwise (\textit{cw}) or counter-clockwise (\textit{ccw}) around the central \hkl[111] direction.
	Visual inspection allows to assign the \textit{ccw} features in pattern G1 in (b) to the simulation for space group P\,$4_132$ in (a), and the \textit{cw} features in pattern G2 in (e) correspond to P\,$4_332$ in (d). A detailed intensity profile analysis of the features is shown in (c) and (f). 
	%In the table, the visual assignment is quantified by numerical values of the normalized cross-correlation coefficient $r$ between the experimental patterns and the respective simulations. 
	%The primary beam energy of the SEM was 20\,kV. 
	For an animated version of this data, see the supplementary  \href{https://github.com/wiai/kikuchi-acentric}{online data repository} \protect\cite{repo_mn}.
	}
	\label{fig:expsim111}
\end{figure}

The polarity and chirality of crystal structures can be analyzed by Kikuchi diffraction based on the quantitative matching to simulations which apply the dynamical theory of electron diffraction \cite{winkelmann2015apl,winkelmann2015um,winkelmann2009ebsd2}.
The breaking of mirror symmetry in the $\beta$-Mn structure is reflected in a corresponding symmetry breaking in the Kikuchi patterns of both possible variants.
This can be seen in Figure\,\ref{fig:expsim111}, where we show experimental Kikuchi pattern data which was measured from two different grains (G1, G2) of a polycrystalline $\beta$-Mn sample. 
Similar to the procedure described in \cite{winkelmann2020amat}, the calibrated experimental patterns measured from differently oriented $\beta$-Mn grains were numerically reprojected and symmetrized according to a three-fold rotation so that the three-fold rotation axis along the \hkl[111] zone axis is at the center of the projection.
In the oval areas marked in the experimental patterns in the middle column of Figure\,\ref{fig:expsim111}, we can note slightly asymmetric features, which allow to define a sense of rotation around the central \hkl[111] direction. 
The arrows in Figure\,\ref{fig:expsim111} point from the lower intensity part of the asymmetric features to the higher intensity part of the two-part feature, and in this way define a counter clockwise (\textit{ccw}) rotation for the upper pattern measured from grain G1, and a clockwise (\textit{cw}) rotation for the lower pattern measured from the second grain G2.
The proper correspondence between the experimentally observed sense of rotation and the specific atomic coordinates in the crystal unit cell (i.e. the absolute structure) can be established by a comparison of the experimental patterns to the theoretical calculations shown in the left column of Figure\,\ref{fig:expsim111}.
This allows to assign the experimental \textit{ccw} pattern G1 to the simulation for space group $P\,4_132$, and the \textit{cw} pattern G2 to $P\,4_332$, respectively.

\begin{figure}[!htbp]
  \centering
  \includegraphics[width=8.5cm]{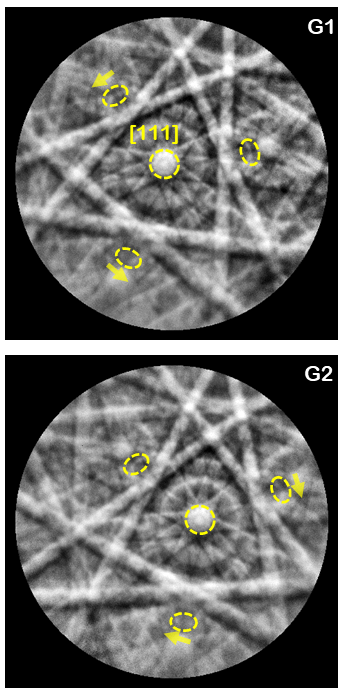}
   \caption{Identification of the sense of rotation around the three-fold \hkl[111] zone axis in experimental Kikuchi patterns from two different $\beta$-Mn grains before the reprojection and three-fold rotational symmetrization for Figure\,\ref{fig:expsim111}  (G1:\,counter-clockwise, G2:\,clockwise). Measurements at 20kV, 480x480 pixels, background processed raw data from a Hikari EBSD detector (EDAX). In both patterns, the sense of rotation can be reliably identified visually in at least two out of three equivalent regions of interest (ROI).The arrows point parallel to the directions from the lower intensity feature to the higher intensity in the respective ROI. The specific space group which corresponds to each of the two patterns is assigned based on the comparison to explicit Kikuchi pattern simulations as discussed in the main text.}
\label{fig:mn111edax}
\end{figure}

The required mirror-relationship between the asymmetric features marked in the Kikuchi patterns of Figure\,\ref{fig:expsim111}(a,b,d,e) can be seen in detail in Figure\,\ref{fig:expsim111}(c) and (f), where we show a comparison of the intensity profiles which were obtained by integration along the vertical direction of the region of interest. 
In agreement with the Kikuchi patterns, the corresponding theoretical profiles in Figure\,\ref{fig:expsim111} show a very good specific agreement of the experimental data from grain G1 to space group $P\,4_132$ and for the grain G2 to space group $P\,4_332$.
In Figure\,\ref{fig:mn111edax}, we also demonstrate the visual identification of the specific enantiomorph from as-measured Kikuchi patterns in different orientations.

\begin{figure}[!htbp]
  \centering
  \includegraphics[width=14.5cm]{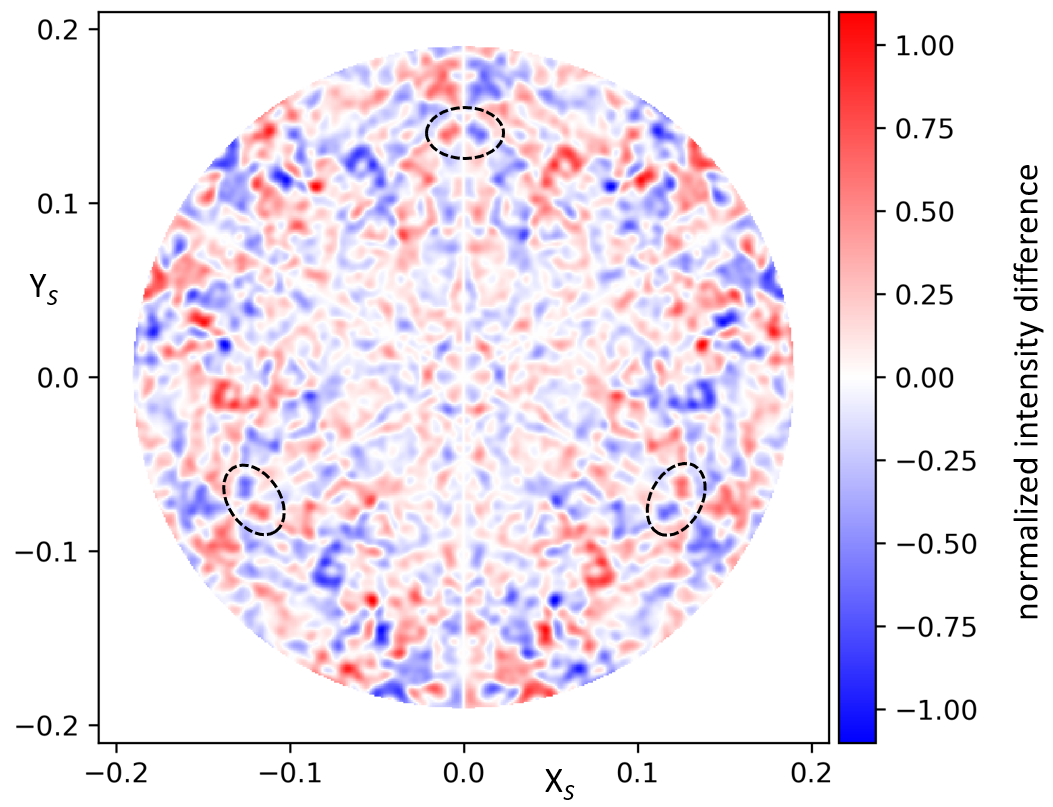}
   \caption{Asymmetry in calculated Kikuchi diffraction patterns   $\beta$-Mn near the three-fold [111] zone axis, in stereographic projection. The asymmetry is determined as the difference between the simulated $\beta$-Mn Kikuchi patterns for space groups $P\,4_132$ and $P\,4_332$ as seen in Figure\,\ref{fig:expsim111}.
   The stereographic projection coordinates $(X_S, Y_S)$ are defined by $\rho = \sqrt{X_S^2 +Y_S^2} = 0.5\tan{\vartheta}$, where $\vartheta$ is the polar angle measured from $(0,0)$. 
   The features which were analyzed in Figure \ref{fig:expsim111} in the main text are marked by ovals. These specific features are particularly suited to visually recognize the breaking of mirror symmetry, because the asymmetry is expressed by two separated features which can be distinguished against a low local intensity background near these features.
   For the quantitative comparison of the experimental patterns with simulations, all of the various difference-features shown here contribute to the change the normalized cross correlation coefficients discussed for Figure\,\ref{fig:expsim111} in the main text. 
   }
\label{fig:asymmetry}
\end{figure}

In addition to the specific key features which we emphasized in Figure\,\ref{fig:expsim111}, there are various other regions in the Kikuchi patterns of $\beta$-Mn which are sensitive to a change in the sense of chirality, but which are harder to identify visually, see Figure\,\ref{fig:asymmetry}. 
The combined systematic contribution of all the chirality-sensitive features can be captured by a quantitative numerical comparison of the experimental and simulated Kikuchi patterns, using, for example, the normalized cross correlation coefficient (NCC) $r$ as an image similarity measure \cite{metcalfe1994book}.
In the following, the coefficients $r_+$ and $r_-$, respectively, will quantitatify the comparison against simulations for space group  $P\,4_132$ and  $P\,4_332$ of $\beta$-Mn.
For the patterns shown in Figure\,\ref{fig:expsim111}, we calculated the values of the NCC resulting from a comparison of each experimental pattern, G1 and G2, against the simulations for the two space groups.
The assignment of the G1 experimental data is quantified by a higher $r_+=0.883$, compared to $r_-=0.871$, while the G2 experiment shows a higher $r_- = 0.873$, compared to $r_+=0.861$. 
The significance of the discrimination between the simulated patterns of the enantiomorphs can be estimated by the value of the difference $\Delta r =  r_+ - r_-$, where quantitative values of $|\Delta r| \gtrapprox 0.01$ can be considered as sufficient for the range of mean NCC values  $r_m =  (r_- + r_+)/2$ which are observed in the present study  \cite{burkhardt2020srep}.
For $\beta$-Mn, we can thus conclude that a reliable assignment of the sense of chirality is possible using Kikuchi diffraction patterns measured in the SEM. Furthermore, this also applies to multi-element compounds with the $\beta$-Mn structure type, for which the sense of chirality can be confirmed by X-ray diffraction using anomalous scattering effects \cite{burkhardt2021sciadv}.

\subsection{EBSD mapping of the local sense of chirality in $\beta$-Mn}
\label{sec:mapping}

\begin{figure}[!htb]
	\centering
	\includegraphics[width=8.8cm]{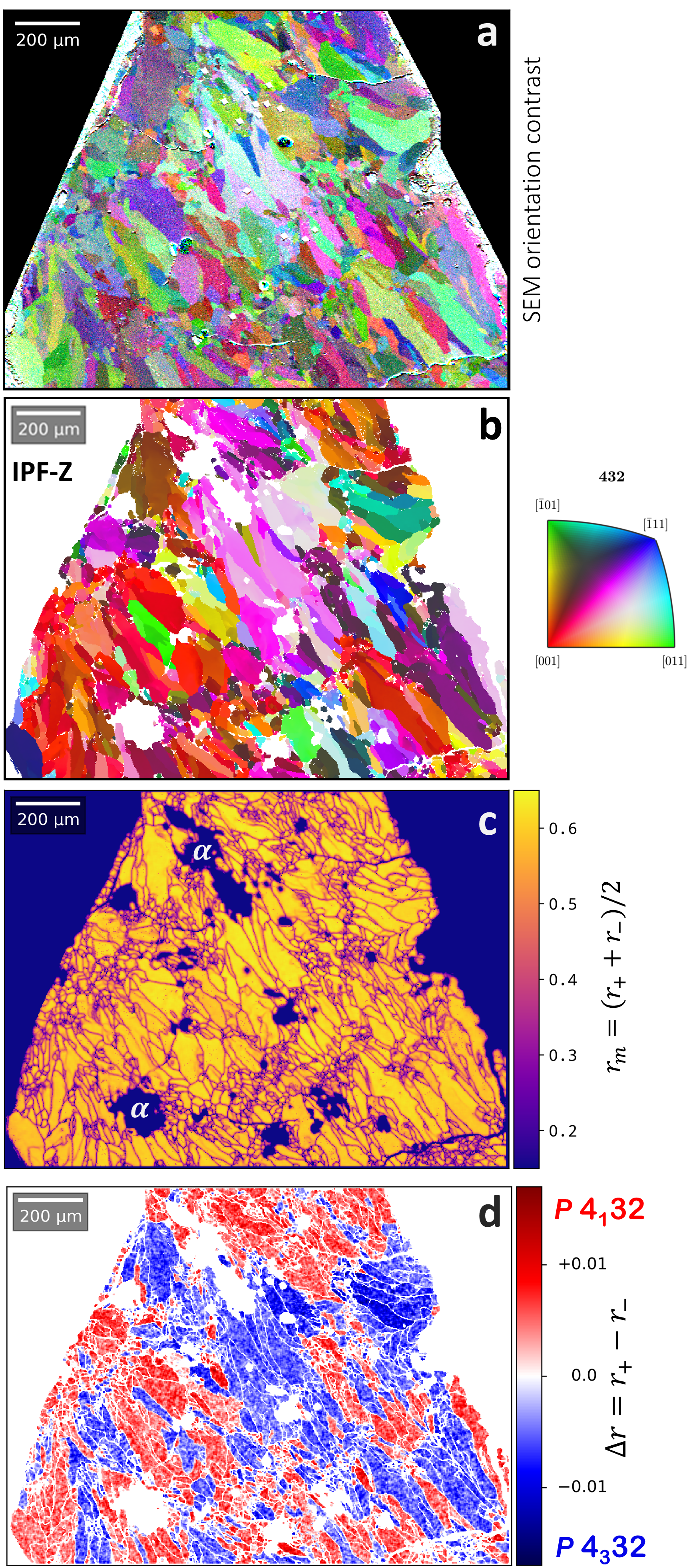}
	\caption{
    	Chirality mapping of polycrystalline $\beta$-Mn sample. (a) qualitative orientation grain contrast image based on the differential backscattered electron signal from the EBSD detector, (b) inverse pole figure map (IPF-Z) of the $\beta$-Mn orientation data,  (c) mean correlation coefficient $r_m = (r_+ - r_-)/2$ relative to simulations for the enantiomorphs of $\beta$-Mn ($r_+ \mathrel{\widehat{=}} P\,4_132$, $r_- \mathrel{\widehat{=}} P\,4_332$), (d) observed difference $r_+ - r_-$ indicating the best-fit space group. Grain boundaries and grains of $\alpha$-Mn are seen as white areas in (d), which have been excluded based on their low values of $r_m<0.4$ in map (c).
    	\label{fig:mnmap}
	}
\end{figure}

Automated quantitative pattern matching provides the basis for the spatially resolved mapping of the chirality of $\beta$-Mn samples in the SEM.
In Figure\,\ref{fig:mnmap}, we show the result of mapping a sample area of \mbox{1.56\,mm $\times$ 1.17\,mm} with a step size of 2.6\,$\mu$m, corresponding to a total number $600\times450$ measured EBSD patterns
with a resolution of $320\times240$ pixels.

In Figure\,\ref{fig:mnmap}(a), the polycrystalline microstructure of a $\beta$-Mn sample is imaged by a qualitative grain orientation contrast map, which was obtained from the measured raw Kikuchi patterns by assigning a color signal to the differential intensity variations measured on different sub-areas of the EBSD detector \cite{winkelmann2018prm}.
Compared to other imaging modes based on the intensity in a set of sub-areas of raw or processed EBSD patterns \cite{wright2015um,brodusch2018joi}, the differential mode assigns the $R,G,B$ color channels to relative intensity changes between specified pairs of sub-areas.
Specifically, for a binning of the original EBSD pattern to an array $I_{jk}$ of $7\times7$ intensities ($j,k=1\ldots7$), we have assigned the color channels to vary according to the ratios of intensities in row 3 and row 2 of the binned array, i.e.  $R \propto I_{32}/I_{22}$, $G \propto I_{34}/I_{24}$, $B \propto I_{36}/I_{26}$.
As can be seen by comparison to the  quantitative orientation data of $\beta$-Mn shown as an inverse pole figure map (IPF-Z) in Figure\,\ref{fig:mnmap}(b), the differential imaging mode leads to color assignments which reveal the grain orientation contrast of the polycrystalline structure directly from the measured intensities, without any explicit Kikuchi pattern orientation analysis
(see \cite{xcdskd_bse} for additional examples and Python code).
We emphasize that the actual grain colors in Figure\,\ref{fig:mnmap}(a) and in Figure\,\ref{fig:mnmap}(b) are not directly related, but in each case, a similar color is assigned to points in the same grain.

In the pattern matching analysis, the best-fit NCC was determined for both possible space groups for each of the Kikuchi patterns of the map, which resulted in the local values of  $r_+$ ($P\,4_132$) and $r_-$ ($P\,4_332$). From these values, the maps of the mean NCC $r_m$ and the difference map $\Delta r$ were derived, which are shown in  Figure\,\ref{fig:mnmap}(c) and (d), respectively.
It needs to be taken into account that the measured data points from the sample can have a varying significance, for example because of lower quality Kikuchi patterns which are observed at grain boundaries, in contaminated or deformed regions, or due to the presence of additional phases.
In order to account for the different reliability of the discrimination between enantiomorphs and to exclude data points which cannot be trusted to indicate a significant chirality dependence, we have not considered data points with $r_m<0.4$ for the map of $\Delta r$ in Figure\,\ref{fig:mnmap}(d).
In the map shown Figure\,\ref{fig:mnmap}(d), the sign of $\Delta r$ indicates the best-fit space group, i.e. positive for  $P\,4_132$ and  negative for $P\,4_332$. 
Absolute values in the order of   $\Delta r \gtrapprox 0.01$ indicate a reliable assignment of the space group and can be seen in Figure\,\ref{fig:mnmap}(d) as red colors for positive $\Delta r$ ($P\,4_132$) and blue colors for negative $\Delta r$ ($P\,4_332$). 
It can be observed that map points within the individual $\beta$-Mn grains are consistently identified with the same space group.
The lighter shades Figure\,\ref{fig:mnmap}(c) indicate unreliable data points at grain boundaries, or grains of $\alpha$-Mn. 
As can be seen by the similar numbers of red and blue grains in Figure\,\ref{fig:mnmap}(d), the distribution of the two enantiomorphs in the investigated sample does not show a significant preference for either of the two enantiomorphs.

We note that the pair of Figures \,\ref{fig:mnmap}(b) and (d) is an example for a generalized orientation description of microstructures which contain chiral or otherwise non-centrosymmetric phases, combining the rotation information (b), with a possible inversion (d) relative to a reference structure \cite{esling1980jdp,bunge1980jac}.

\FloatBarrier
\subsection{Chirality discrimination of Tellurium and Selenium}

As an additional application of the quantitative pattern matching approach to chirality discrimination of elemental crystals, we also analyze Kikuchi diffraction patterns of trigonal tellurium and selenium. 

The chirality of tellurium so far has been analyzed using polarized neutron scattering \cite{brown1996aca} and by X-ray diffraction using circularly polarized synchrotron radiation \cite{tanaka2008prl,tanaka2010jpcm}. 
Chirality specification of macroscopic Te samples has also been discussed by the characterization of etch-pits on surfaces of single crystals \cite{sakano2020prl}, which can be grown by a physical vapor transport technique \cite{ideue2019pnas}.

\begin{figure}[!htb]
	\centering
	\includegraphics[width=7.0cm]{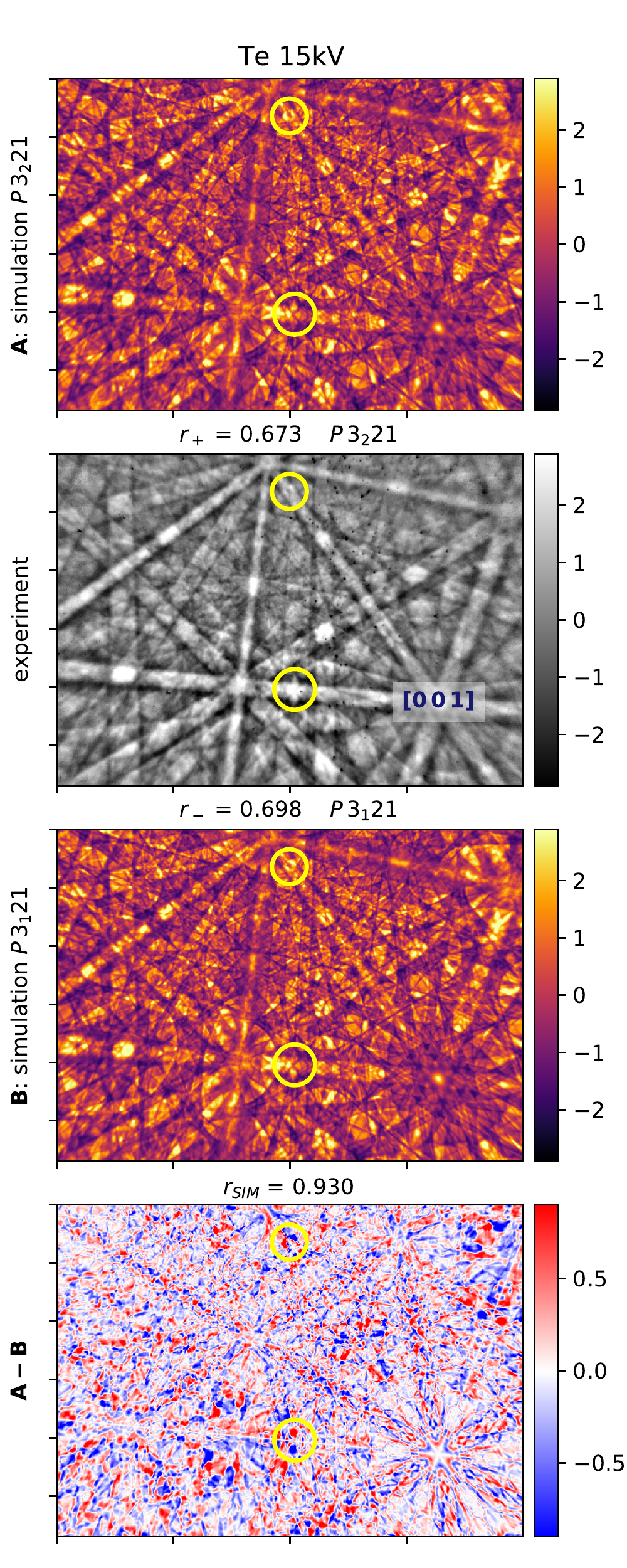}
	\includegraphics[width=7.0cm]{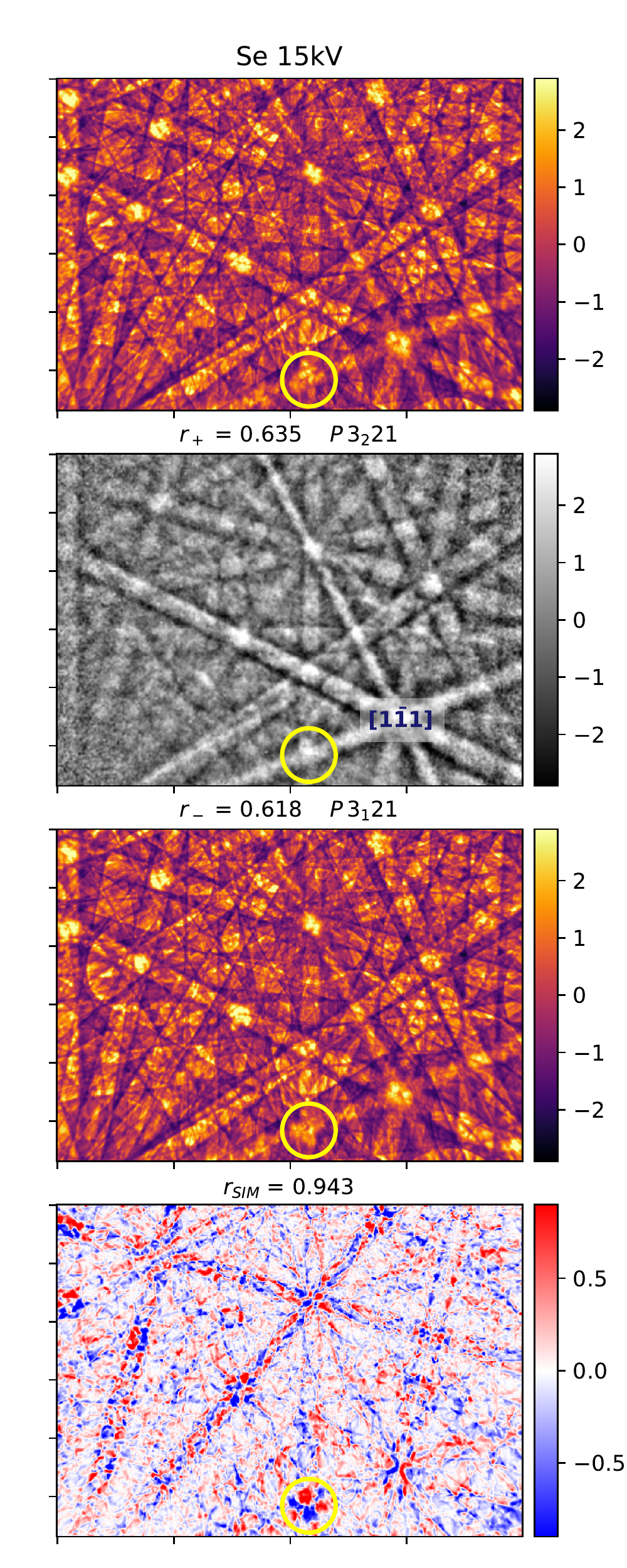}
	\caption{Discrimination of the sense of chirality in trigonal tellurium (left column) and selenium (right column) using Kikuchi diffraction patterns at 15kV. 
	The experimental patterns (second row) are compared to simulations for both possible enantiomorphs (top row: \mbox{A = $P\,3_221$}; third row: \mbox{B = $P\,3_121$}). 
	Bottom row: difference of the simulations for both variants, showing the distribution of chirality-sensitive features in the complete pattern.
	For visual comparison, some characteristic features are marked by circles.
	For the investigated Te sample, the larger  value of the normalized cross correlation coefficient $r_-=0.698$ indicates space group $P\,3_121$, with $|\Delta r| = 0.025$.
	For the investigated Se sample, the larger  value of the normalized cross correlation coefficient $r_+=0.635$ indicates space group $P\,3_221$, with  $|\Delta r| = 0.017$. 
	The Se simulations show a similar theoretical sensitivity to chirality as the Te patterns, with the the correlation between the simulated variants $r_{SIM}=0.93$ and $0.94$ for Te and Se, respectively.
    (for animated figures, see also \cite{repo_mn})    
	\label{fig:Te15kVpattern}}
\end{figure}

In the left column of Figure\,\ref{fig:Te15kVpattern}, we show the comparison of an experimental Kikuchi pattern measured from a Te crystallite at 15\,kV  (second row) to best-fit simulations for the two cases of space group $P\,3_221$  (A, top row, $r_+$) and space group $P\,3_121$ (B, third row, $r_-$), respectively. 
For the Kikuchi diffraction simulations, we have used the Te crystal structure definition for space group $P\,3_121$ with $a=4.4572$\AA, $c=5.929$\AA, and the Te $3a$ atomic position at $(0.2633, 0, 1/3)$ \cite{cherin1967aca}.

As can be seen in the left column of Figure\,\ref{fig:Te15kVpattern}, the Te simulations show a very good overall agreement with the experiment, which is quantified by the respective normalized cross correlation coefficients $r_+=0.673$ and $r_-=0.698$. 
The larger value of $r_-=0.698$ corresponds to the simulated pattern of space group $P\,3_121$, with a significant difference to the alternative space group $P\,3_221$ by $|\Delta r| = 0.025$.
In the bottom row of Figure\,\ref{fig:Te15kVpattern}, the difference of the simulations for both variants shows the distribution of chirality-sensitive features in the Kikuchi pattern. 
Some characteristic features are marked by circles in Figure\,\ref{fig:Te15kVpattern}, in order to enable also a visual verification of the chirality effects in the patterns.

In in the right column of Figure\,\ref{fig:Te15kVpattern}, we show a similar analysis for a Kikuchi pattern from trigonal selenium, which has the same structure type as tellurium. 
The Se crystal structure definition for space group $P\,3_121$ has $a=4.3662$\AA, $c=4.9536$\AA, with the Se $3a$ atomic position at $(0.2254, 0, 1/3)$ \cite{cherin1967ioc}.
For the investigated Se sample, the larger  value of the normalized cross correlation coefficient $r_+=0.635$ indicates space group $P\,3_221$, with  $|\Delta r| = 0.017$.

For Te and Se, we expect that near the \hkl(001) plane, the chirality effects are absent due to the two-fold rotation axis which is contained in this plane, the corresponding \hkl(001), however is not present for the orienation of the measured crystal relative to the detector.  
A comparison of the results for Te and Se in  Figure\,\ref{fig:Te15kVpattern} shows that the chirality-sensitive differences in the Kikuchi diffraction patterns of both elements are of a similar magnitude. 
This is also expressed by the similar correlation coefficients between the two ideal, theoretical enantiomorph patterns, with $r_{SIM}=0.930$ for the two simulated Te patterns compared to $r_{SIM}=0.943$ for the simulated Se patterns.
The clear discrimination of the sense of chirality of the Te and Se in Figure\,\ref{fig:Te15kVpattern} indicates that it should be possible to use electron backscatter diffraction to analyze the local chirality of Te and Se materials in a laboratory SEM in a similar way as discussed above for $\beta$-Mn.

\section{Summary \& Conclusion}

Kikuchi electron diffraction patterns allow to assign the sense of chirality for elemental crystals of cubic $\beta$-Manganese, as well as for trigonal tellurium and selenium.
Compared to more demanding measurements in the transmission electron microscope \cite{tanaka1985jpsj,ma2017nmat,dong2020ncomm,brazda2019science}, synchrotron-based polarized X-ray diffraction \cite{tanaka2008prl,tanaka2010jpcm,morikawa2020prm,lovesey2021prb}, or polarized neutron diffraction \cite{brown1996aca},
the method discussed here can be implemented relatively easily in laboratory scanning electron microscopes. 
In the SEM, chirality-sensitive Kikuchi diffraction patterns can be measured by the methods of EBSD and TKD, with reduced requirements on sample preparation and experimental complexity, and the additional possibility to quickly investigate the local sense of chirality in large sample areas at high spatial resolution down to the range of a few tens of nanometers.
In this way, our study provides an efficient method for the analysis of chirality-dependend effects in elemental materials on the spatial scales which are relevant for potential applications.

\bigskip
%%%%%%%%%%%%%%%%%%%%%%%%%%%%%%%%%%%%%%%%%%

%%%%%%%%%%%%%%%%%%%%%%%%%%%%%%%%%%%%%%%%%%
\funding{This research was supported by the Polish National Agency for Academic Exchange (NAWA) grant numbers PPI/APM/2018/1/00049/U/001 and PPN/ULM/2019/1/00068/U/00001, as well as by the Polish National Science Centre (NCN) grant number \mbox{2020/37/B/ST5/03669}.}

%%%%%%%%%%%%%%%%%%%%%%%%%%%%%%%%%%%%%%%%%%
\acknowledgments{We thank Andreea Dumitriu and Dr. Marcus Schmidt (
Max-Planck-Institut für Chemische Physik fester Stoffe, Dresden) for the preparation of Mn, Te, and Se crystals.} 

%%%%%%%%%%%%%%%%%%%%%%%%%%%%%%%%%%%%%%%%%%
\conflictsofinterest{The authors declare no conflict of interest.} 
%%%%%%%%%%%%%%%%%%%%%%%%%%%%%%%%%%%%%%%%%%

%\bibliography{awlibrary}

\begin{thebibliography}{-------}
\providecommand{\natexlab}[1]{#1}

\bibitem[Flack(2003)]{flack2003hca}
Flack, H.D.
\newblock Chiral and Achiral Crystal Structures.
\newblock {\em Helvetica Chimica Acta} {\bf 2003}, {\em 86},~905--921.
\newblock
  doi:{\changeurlcolor{black}\href{https://doi.org/10.1002/hlca.200390109}{\detokenize{10.1002/hlca.200390109}}}.

\bibitem[Shoemaker \em{et~al.}(1978)Shoemaker, Shoemaker, Hopkins, and
  Yindepit]{shoemaker1978acb}
Shoemaker, C.B.; Shoemaker, D.P.; Hopkins, T.E.; Yindepit, S.
\newblock Refinement of the structure of $\beta$-{M}anganese and of a related
  phase in the {Mn-Ni-Si} system.
\newblock {\em Acta Cryst. B} {\bf 1978}, {\em 34},~3573--3576.
\newblock
  doi:{\changeurlcolor{black}\href{https://doi.org/10.1107/s0567740878011620}{\detokenize{10.1107/s0567740878011620}}}.

\bibitem[Cherin and Unger(1967{\natexlab{a}})]{cherin1967aca}
Cherin, P.; Unger, P.
\newblock Two-dimensional refinement of the crystal structure of tellurium.
\newblock {\em Acta Crystallographica} {\bf 1967}, {\em 23},~670--671.
\newblock
  doi:{\changeurlcolor{black}\href{https://doi.org/10.1107/s0365110x6700341x}{\detokenize{10.1107/s0365110x6700341x}}}.

\bibitem[Cherin and Unger(1967{\natexlab{b}})]{cherin1967ioc}
Cherin, P.; Unger, P.
\newblock The crystal structure of trigonal selenium.
\newblock {\em Inorganic Chemistry} {\bf 1967}, {\em 6},~1589--1591.
\newblock
  doi:{\changeurlcolor{black}\href{https://doi.org/10.1021/ic50054a037}{\detokenize{10.1021/ic50054a037}}}.

\bibitem[Newnham(2005)]{newnham2005properties}
Newnham, R.
\newblock {\em Properties of materials: anisotropy, symmetry, structure};
  Oxford University Press: Oxford New York,  2005.

\bibitem[Barron(2009)]{barron2009chapter}
Barron, L.D.
\newblock An Introduction to Chirality at the Nanoscale. In {\em Chirality at
  the Nanoscale: Nanoparticles, Surfaces, Materials and more.}; Amabilino,
  D.B., Ed.; WILEY-VCH: Weinheim,  2009; pp. 1--27.
\newblock
  doi:{\changeurlcolor{black}\href{https://doi.org/10.1002/9783527625345.ch1}{\detokenize{10.1002/9783527625345.ch1}}}.

\bibitem[Hafner and Hobbs(2003)]{hafner2003prb2}
Hafner, J.; Hobbs, D.
\newblock Understanding the complex metallic element {Mn}. \,{II}. {G}eometric
  frustration in $\beta$-{Mn}, phase stability, and phase transitions.
\newblock {\em Physical Review B} {\bf 2003}, {\em 68}.
\newblock
  doi:{\changeurlcolor{black}\href{https://doi.org/10.1103/physrevb.68.014408}{\detokenize{10.1103/physrevb.68.014408}}}.

\bibitem[Preston(1928)]{preston1928pma}
Preston, G.D.
\newblock The crystal structure of $\beta$-{Manganese}.
\newblock {\em The London, Edinburgh, and Dublin Philosophical Magazine and
  Journal of Science} {\bf 1928}, {\em 5},~1207--1225.
\newblock
  doi:{\changeurlcolor{black}\href{https://doi.org/10.1080/14786440608564570}{\detokenize{10.1080/14786440608564570}}}.

\bibitem[{Pickworth Glusker} and Trueblood(2010)]{glusker2010}
{Pickworth Glusker}, J.; Trueblood, K.N.
\newblock {\em {Crystal Structure Analysis : A Primer}}; Oxford University
  Press: Oxford New York,  2010.

\bibitem[Flack and Bernardinelli(2008)]{flack2008chirality}
Flack, H.D.; Bernardinelli, G.
\newblock {The use of X-ray crystallography to determine absolute
  configuration}.
\newblock {\em Chirality} {\bf 2008}, {\em 20},~681--690.
\newblock
  doi:{\changeurlcolor{black}\href{https://doi.org/10.1002/chir.20473}{\detokenize{10.1002/chir.20473}}}.

\bibitem[Tanaka \em{et~al.}(2010)Tanaka, Collins, Lovesey, Matsumami, Moriwaki,
  and Shin]{tanaka2010jpcm}
Tanaka, Y.; Collins, S.P.; Lovesey, S.W.; Matsumami, M.; Moriwaki, T.; Shin, S.
\newblock {Determination of the absolute chirality of tellurium using resonant
  diffraction with circularly polarized x-rays.}
\newblock {\em J. Phys. Condens. Matter} {\bf 2010}, {\em 22},~122201.
\newblock
  doi:{\changeurlcolor{black}\href{https://doi.org/10.1088/0953-8984/22/12/122201}{\detokenize{10.1088/0953-8984/22/12/122201}}}.

\bibitem[Burkhardt \em{et~al.}(2021)Burkhardt, Winkelmann, Borrmann, Dumitriu,
  König, Cios, and Grin]{burkhardt2021sciadv}
Burkhardt, U.; Winkelmann, A.; Borrmann, H.; Dumitriu, A.; König, M.; Cios,
  G.; Grin, Y.
\newblock {A}ssignment of enantiomorphs for the chiral allotrope $\beta$-{Mn}
  by diffraction methods.
\newblock {\em Science Advances} {\bf 2021}, {\em 7},~eabg0868.
\newblock
  doi:{\changeurlcolor{black}\href{https://doi.org/10.1126/sciadv.abg0868}{\detokenize{10.1126/sciadv.abg0868}}}.

\bibitem[Nakamura and Shiga(1997)]{nakamura1997pb}
Nakamura, H.; Shiga, M.
\newblock Frustration in $\beta$-{Mn}.
\newblock {\em Physica B: Condensed Matter} {\bf 1997}, {\em
  237-238},~453--454.
\newblock
  doi:{\changeurlcolor{black}\href{https://doi.org/10.1016/s0921-4526(97)00129-4}{\detokenize{10.1016/s0921-4526(97)00129-4}}}.

\bibitem[Eriksson \em{et~al.}(2005)Eriksson, Bergqvist, Andersson, Nordblad,
  and Eriksson]{eriksson2005prb}
Eriksson, T.; Bergqvist, L.; Andersson, Y.; Nordblad, P.; Eriksson, O.
\newblock Magnetic properties of selected {Mn}-based transition metal compounds
  with $\beta$-{Mn} structure: {E}xperiments and {T}heory.
\newblock {\em Physical Review B} {\bf 2005}, {\em 72},~144427.
\newblock
  doi:{\changeurlcolor{black}\href{https://doi.org/10.1103/physrevb.72.144427}{\detokenize{10.1103/physrevb.72.144427}}}.

\bibitem[Paddison \em{et~al.}(2013)Paddison, Stewart, Manuel, Courtois,
  McIntyre, Rainford, and Goodwin]{paddison2013prl}
Paddison, J.A.M.; Stewart, J.R.; Manuel, P.; Courtois, P.; McIntyre, G.J.;
  Rainford, B.D.; Goodwin, A.L.
\newblock {Emergent Frustration in Co-doped $\beta$-Mn}.
\newblock {\em Physical Review Letters} {\bf 2013}, {\em 110}.
\newblock
  doi:{\changeurlcolor{black}\href{https://doi.org/10.1103/physrevlett.110.267207}{\detokenize{10.1103/physrevlett.110.267207}}}.

\bibitem[Tokunaga \em{et~al.}(2015)Tokunaga, Yu, White, R{\o}nnow, Morikawa,
  Taguchi, and Tokura]{tokunaga2015ncomm}
Tokunaga, Y.; Yu, X.Z.; White, J.S.; R{\o}nnow, H.M.; Morikawa, D.; Taguchi,
  Y.; Tokura, Y.
\newblock A new class of chiral materials hosting magnetic skyrmions beyond
  room temperature.
\newblock {\em Nature Communications} {\bf 2015}, {\em 6},~7638.
\newblock
  doi:{\changeurlcolor{black}\href{https://doi.org/10.1038/ncomms8638}{\detokenize{10.1038/ncomms8638}}}.

\bibitem[Rö{\ss}ler \em{et~al.}(2006)Rö{\ss}ler, Bogdanov, and
  Pfleiderer]{roesler2006nat}
Rö{\ss}ler, U.K.; Bogdanov, A.N.; Pfleiderer, C.
\newblock Spontaneous skyrmion ground states in magnetic metals.
\newblock {\em Nature} {\bf 2006}, {\em 442},~797--801.
\newblock
  doi:{\changeurlcolor{black}\href{https://doi.org/10.1038/nature05056}{\detokenize{10.1038/nature05056}}}.

\bibitem[Nagaosa and Tokura(2013)]{nagaosa2013nnano}
Nagaosa, N.; Tokura, Y.
\newblock Topological properties and dynamics of magnetic skyrmions.
\newblock {\em Nature Nanotechnology} {\bf 2013}, {\em 8},~899--911.
\newblock
  doi:{\changeurlcolor{black}\href{https://doi.org/10.1038/nnano.2013.243}{\detokenize{10.1038/nnano.2013.243}}}.

\bibitem[Karube \em{et~al.}(2018)Karube, Shibata, White, Koretsune, Yu,
  Tokunaga, R\o{}nnow, Arita, Arima, Tokura, and Taguchi]{karube2018prb}
Karube, K.; Shibata, K.; White, J.S.; Koretsune, T.; Yu, X.Z.; Tokunaga, Y.;
  R\o{}nnow, H.M.; Arita, R.; Arima, T.; Tokura, Y.; Taguchi, Y.
\newblock Controlling the helicity of magnetic skyrmions in a
  $\ensuremath{\beta}$-{Mn}-type high-temperature chiral magnet.
\newblock {\em Phys. Rev. B} {\bf 2018}, {\em 98},~155120.
\newblock
  doi:{\changeurlcolor{black}\href{https://doi.org/10.1103/PhysRevB.98.155120}{\detokenize{10.1103/PhysRevB.98.155120}}}.

\bibitem[Chang \em{et~al.}(2018)Chang, Wieder, Schindler, Sanchez, Belopolski,
  Huang, Singh, Wu, Chang, Neupert, Xu, Lin, and Hasan]{chang2018nmat}
Chang, G.; Wieder, B.J.; Schindler, F.; Sanchez, D.S.; Belopolski, I.; Huang,
  S.M.; Singh, B.; Wu, D.; Chang, T.R.; Neupert, T.; Xu, S.Y.; Lin, H.; Hasan,
  M.Z.
\newblock Topological quantum properties of chiral crystals.
\newblock {\em Nature Materials} {\bf 2018}, {\em 17},~978--985.
\newblock
  doi:{\changeurlcolor{black}\href{https://doi.org/10.1038/s41563-018-0169-3}{\detokenize{10.1038/s41563-018-0169-3}}}.

\bibitem[Tsirkin \em{et~al.}(2018)Tsirkin, Puente, and Souza]{tsirkin2018prb}
Tsirkin, S.S.; Puente, P.A.; Souza, I.
\newblock Gyrotropic effects in trigonal tellurium studied from first
  principles.
\newblock {\em Physical Review B} {\bf 2018}, {\em 97}.
\newblock
  doi:{\changeurlcolor{black}\href{https://doi.org/10.1103/physrevb.97.035158}{\detokenize{10.1103/physrevb.97.035158}}}.

\bibitem[Ideue \em{et~al.}(2019)Ideue, Hirayama, Taiko, Takahashi, Murase,
  Miyake, Murakami, Sasagawa, and Iwasa]{ideue2019pnas}
Ideue, T.; Hirayama, M.; Taiko, H.; Takahashi, T.; Murase, M.; Miyake, T.;
  Murakami, S.; Sasagawa, T.; Iwasa, Y.
\newblock Pressure-induced topological phase transition in noncentrosymmetric
  elemental tellurium.
\newblock {\em {Proceedings of the National Academy of Sciences}} {\bf 2019},
  {\em 116},~25530--25534.
\newblock
  doi:{\changeurlcolor{black}\href{https://doi.org/10.1073/pnas.1905524116}{\detokenize{10.1073/pnas.1905524116}}}.

\bibitem[Sakano \em{et~al.}(2020)Sakano, Hirayama, Takahashi, Akebi, Nakayama,
  Kuroda, Taguchi, Yoshikawa, Miyamoto, Okuda, Ono, Kumigashira, Ideue, Iwasa,
  Mitsuishi, Ishizaka, Shin, Miyake, Murakami, Sasagawa, and
  Kondo]{sakano2020prl}
Sakano, M.; Hirayama, M.; Takahashi, T.; Akebi, S.; Nakayama, M.; Kuroda, K.;
  Taguchi, K.; Yoshikawa, T.; Miyamoto, K.; Okuda, T.; Ono, K.; Kumigashira,
  H.; Ideue, T.; Iwasa, Y.; Mitsuishi, N.; Ishizaka, K.; Shin, S.; Miyake, T.;
  Murakami, S.; Sasagawa, T.; Kondo, T.
\newblock {Radial Spin Texture in Elemental Tellurium with Chiral Crystal
  Structure}.
\newblock {\em Physical Review Letters} {\bf 2020}, {\em 124},~136404.
\newblock
  doi:{\changeurlcolor{black}\href{https://doi.org/10.1103/PhysRevLett.124.136404}{\detokenize{10.1103/PhysRevLett.124.136404}}}.

\bibitem[Tanaka \em{et~al.}(2008)Tanaka, Takeuchi, Lovesey, Knight, Chainani,
  Takata, Oura, Senba, Ohashi, and Shin]{tanaka2008prl}
Tanaka, Y.; Takeuchi, T.; Lovesey, S.W.; Knight, K.S.; Chainani, A.; Takata,
  Y.; Oura, M.; Senba, Y.; Ohashi, H.; Shin, S.
\newblock {Right Handed or Left Handed? Forbidden X-Ray Diffraction Reveals
  Chirality}.
\newblock {\em Physical Review Letters} {\bf 2008}, {\em 100}.
\newblock
  doi:{\changeurlcolor{black}\href{https://doi.org/10.1103/physrevlett.100.145502}{\detokenize{10.1103/physrevlett.100.145502}}}.

\bibitem[Brown and Forsyth(1996)]{brown1996aca}
Brown, P.J.; Forsyth, J.B.
\newblock {The Crystal Structure and Optical Activity of Tellurium}.
\newblock {\em Acta Crystallographica Section A} {\bf 1996}, {\em
  52},~408--412.
\newblock
  doi:{\changeurlcolor{black}\href{https://doi.org/10.1107/S0108767395017144}{\detokenize{10.1107/S0108767395017144}}}.

\bibitem[Lovesey(2021)]{lovesey2021prb}
Lovesey, S.W.
\newblock Structural chirality of $\upbeta$ -{Mn}.
\newblock {\em Physical Review B} {\bf 2021}, {\em 103},~155136.
\newblock
  doi:{\changeurlcolor{black}\href{https://doi.org/10.1103/physrevb.103.155136}{\detokenize{10.1103/physrevb.103.155136}}}.

\bibitem[Goodman and Secomb(1977)]{goodmansecomb1977aca}
Goodman, P.; Secomb, T.W.
\newblock {Identification of enantiomorphously related space groups by electron
  diffraction}.
\newblock {\em Acta Crystallographica Section A} {\bf 1977}, {\em
  33},~126--133.
\newblock
  doi:{\changeurlcolor{black}\href{https://doi.org/10.1107/S0567739477000266}{\detokenize{10.1107/S0567739477000266}}}.

\bibitem[Spence \em{et~al.}(1994)Spence, Zuo, {O'Keeffe}, Marthinsen, and
  Hoier]{spence1994aca}
Spence, J.C.H.; Zuo, J.M.; {O'Keeffe}, M.; Marthinsen, K.; Hoier, R.
\newblock {On the minimum number of beams needed to distinguish enantiomorphs
  in X-ray and electron diffraction}.
\newblock {\em Acta Cryst. A} {\bf 1994}, {\em 50},~647--650.
\newblock
  doi:{\changeurlcolor{black}\href{https://doi.org/10.1107/s0108767394002850}{\detokenize{10.1107/s0108767394002850}}}.

\bibitem[Tanaka \em{et~al.}(1985)Tanaka, Takayoshi, Ishida, and
  Endoh]{tanaka1985jpsj}
Tanaka, M.; Takayoshi, H.; Ishida, M.; Endoh, Y.
\newblock {Crystal Chirality and Helicity of the Helical Spin Density Wave in
  {MnSi}. I.Convergent-Beam Electron Diffraction}.
\newblock {\em Journal of the Physical Society of Japan} {\bf 1985}, {\em
  54},~2970--2974.
\newblock
  doi:{\changeurlcolor{black}\href{https://doi.org/10.1143/JPSJ.54.2970}{\detokenize{10.1143/JPSJ.54.2970}}}.

\bibitem[{Ma} \em{et~al.}(2017){Ma}, {Oleynikov}, and {Terasaki}]{ma2017nmat}
{Ma}, Y.; {Oleynikov}, P.; {Terasaki}, O.
\newblock {Electron crystallography for determining the handedness of a chiral
  zeolite nanocrystal}.
\newblock {\em Nature Materials} {\bf 2017}, {\em 16},~755--759.
\newblock
  doi:{\changeurlcolor{black}\href{https://doi.org/10.1038/nmat4890}{\detokenize{10.1038/nmat4890}}}.

\bibitem[Dong and Ma(2020)]{dong2020ncomm}
Dong, Z.; Ma, Y.
\newblock Atomic-level handedness determination of chiral crystals using
  aberration-corrected scanning transmission electron microscopy.
\newblock {\em Nature Communications} {\bf 2020}, {\em 11},~1588.
\newblock
  doi:{\changeurlcolor{black}\href{https://doi.org/10.1038/s41467-020-15388-5}{\detokenize{10.1038/s41467-020-15388-5}}}.

\bibitem[{Br{\'a}zda} \em{et~al.}(2019){Br{\'a}zda}, {Palatinus}, and
  {Babor}]{brazda2019science}
{Br{\'a}zda}, P.; {Palatinus}, L.; {Babor}, M.
\newblock {Electron diffraction determines molecular absolute configuration in
  a pharmaceutical nanocrystal}.
\newblock {\em Science} {\bf 2019}, {\em 364},~667--669.
\newblock
  doi:{\changeurlcolor{black}\href{https://doi.org/10.1126/science.aaw2560}{\detokenize{10.1126/science.aaw2560}}}.

\bibitem[Schwartz \em{et~al.}(2009)Schwartz, Kumar, Adams, and
  Field]{schwartz2009ebsd2}
Schwartz, A.J.; Kumar, M.; Adams, B.L.; Field, D.P., Eds.
\newblock {\em Electron Backscatter Diffraction in Materials Science}; Springer
  {US},  2009.
\newblock
  doi:{\changeurlcolor{black}\href{https://doi.org/10.1007/978-0-387-88136-2}{\detokenize{10.1007/978-0-387-88136-2}}}.

\bibitem[Baba-Kishi and Dingley(1989)]{babakishi1989sc}
Baba-Kishi, K.Z.; Dingley, D.J.
\newblock {Backscatter Kikuchi diffraction in the SEM for identification of
  crystallographic point groups}.
\newblock {\em Scanning} {\bf 1989}, {\em 11},~305--312.
\newblock
  doi:{\changeurlcolor{black}\href{https://doi.org/10.1002/sca.4950110605}{\detokenize{10.1002/sca.4950110605}}}.

\bibitem[Winkelmann and Nolze(2015)]{winkelmann2015apl}
Winkelmann, A.; Nolze, G.
\newblock {Point-group sensitive orientation mapping of non-centrosymmetric
  crystals}.
\newblock {\em Appl. Phys. Lett.} {\bf 2015}, {\em 106},~072101.
\newblock
  doi:{\changeurlcolor{black}\href{https://doi.org/10.1063/1.4907938}{\detokenize{10.1063/1.4907938}}}.

\bibitem[Sneddon \em{et~al.}(2016)Sneddon, Trimby, and Cairney]{sneddon2016mse}
Sneddon, G.C.; Trimby, P.W.; Cairney, J.M.
\newblock {Transmission Kikuchi diffraction in a scanning electron microscope:
  A Review}.
\newblock {\em Materials Science and Engineering: R: Reports} {\bf 2016}, {\em
  110},~1 -- 12.
\newblock
  doi:{\changeurlcolor{black}\href{https://doi.org/10.1016/j.mser.2016.10.001}{\detokenize{10.1016/j.mser.2016.10.001}}}.

\bibitem[Winkelmann \em{et~al.}(2016)Winkelmann, Nolze, Vos, Salvat-Pujol, and
  Werner]{winkelmann2016iop}
Winkelmann, A.; Nolze, G.; Vos, M.; Salvat-Pujol, F.; Werner, W.S.M.
\newblock {Physics-based simulation models for EBSD: advances and challenges}.
\newblock {\em IOP Conference Series: Materials Science and Engineering} {\bf
  2016}, {\em 109},~012018.
\newblock
  doi:{\changeurlcolor{black}\href{https://doi.org/10.1088/1757-899X/109/1/012018}{\detokenize{10.1088/1757-899X/109/1/012018}}}.

\bibitem[Nolze \em{et~al.}(2015)Nolze, Grosse, and Winkelmann]{nolze2015jac}
Nolze, G.; Grosse, C.; Winkelmann, A.
\newblock Kikuchi pattern analysis of noncentrosymmetric crystals.
\newblock {\em Journal of Applied Crystallography} {\bf 2015}, {\em
  48},~1405--1419.
\newblock
  doi:{\changeurlcolor{black}\href{https://doi.org/10.1107/s1600576715014016}{\detokenize{10.1107/s1600576715014016}}}.

\bibitem[Winkelmann and Nolze(2015)]{winkelmann2015um}
Winkelmann, A.; Nolze, G.
\newblock {Chirality determination of quartz crystals using Electron
  Backscatter Diffraction}.
\newblock {\em Ultramicroscopy} {\bf 2015}, {\em 149},~58--63.
\newblock
  doi:{\changeurlcolor{black}\href{https://doi.org/10.1016/j.ultramic.2014.11.013}{\detokenize{10.1016/j.ultramic.2014.11.013}}}.

\bibitem[Burch \em{et~al.}(2017)Burch, Fancher, Patala, {De Graef}, and
  Dickey]{burch2017um}
Burch, M.J.; Fancher, C.M.; Patala, S.; {De Graef}, M.; Dickey, E.C.
\newblock Mapping 180{\textdegree} polar domains using electron backscatter
  diffraction and dynamical scattering simulations.
\newblock {\em Ultramicroscopy} {\bf 2017}, {\em 173},~47--51.
\newblock
  doi:{\changeurlcolor{black}\href{https://doi.org/10.1016/j.ultramic.2016.11.013}{\detokenize{10.1016/j.ultramic.2016.11.013}}}.

\bibitem[Burkhardt \em{et~al.}(2020)Burkhardt, Borrmann, Moll, Schmidt, Grin,
  and Winkelmann]{burkhardt2020srep}
Burkhardt, U.; Borrmann, H.; Moll, P.; Schmidt, M.; Grin, Y.; Winkelmann, A.
\newblock Absolute Structure from Scanning Electron Microscopy.
\newblock {\em Scientific Reports} {\bf 2020}, {\em 10},~4065.
\newblock
  doi:{\changeurlcolor{black}\href{https://doi.org/10.1038/s41598-020-59854-y}{\detokenize{10.1038/s41598-020-59854-y}}}.

\bibitem[Schäfer and Brendel(1991)]{schaefer1991zaac}
Schäfer, H.; Brendel, C.
\newblock Die Beeinflussung des Kristallhabitus beim chemischen Transport.
\newblock {\em Zeitschrift für anorganische und allgemeine Chemie} {\bf 1991},
  {\em 598},~293--298.
\newblock
  doi:{\changeurlcolor{black}\href{https://doi.org/10.1002/zaac.19915980126}{\detokenize{10.1002/zaac.19915980126}}}.

\bibitem[Nolze \em{et~al.}(2018)Nolze, J{\"u}rgens, Olbricht, and
  Winkelmann]{nolze2018amat}
Nolze, G.; J{\"u}rgens, M.; Olbricht, J.; Winkelmann, A.
\newblock Improving the precision of orientation measurements from technical
  materials via {EBSD} pattern matching.
\newblock {\em Acta Materialia} {\bf 2018}, {\em 159},~408 -- 415.
\newblock
  doi:{\changeurlcolor{black}\href{https://doi.org/10.1016/j.actamat.2018.08.028}{\detokenize{10.1016/j.actamat.2018.08.028}}}.

\bibitem[Winkelmann \em{et~al.}(2020{\natexlab{a}})Winkelmann, Jablon, Tong,
  Trager-Cowan, and Mingard]{winkelmann2020jm}
Winkelmann, A.; Jablon, B.M.; Tong, V.S.; Trager-Cowan, C.; Mingard, K.P.
\newblock Improving {EBSD} precision by orientation refinement with full
  pattern matching.
\newblock {\em Journal of Microscopy} {\bf 2020}, {\em 277},~79--92.
\newblock
  doi:{\changeurlcolor{black}\href{https://doi.org/10.1111/jmi.12870}{\detokenize{10.1111/jmi.12870}}}.

\bibitem[Winkelmann \em{et~al.}(2020{\natexlab{b}})Winkelmann, Nolze, Cios,
  Tokarski, and Ba{\l}a]{winkelmann2020mat}
Winkelmann, A.; Nolze, G.; Cios, G.; Tokarski, T.; Ba{\l}a, P.
\newblock Refined Calibration Model for Improving the Orientation Precision of
  Electron Backscatter Diffraction Maps.
\newblock {\em Materials} {\bf 2020}, {\em 13},~2816.
\newblock
  doi:{\changeurlcolor{black}\href{https://doi.org/10.3390/ma13122816}{\detokenize{10.3390/ma13122816}}}.

\bibitem[Winkelmann(2009)]{winkelmann2009ebsd2}
Winkelmann, A.
\newblock {Dynamical Simulation of Electron Backscatter Diffraction Patterns}.
  In {\em Electron Backscatter Diffraction in Materials Science}; Schwartz,
  A.J.; Kumar, M.; Adams, B.L.; Field, D.P., Eds.; Springer: Berlin,  2009;
  chapter~2, pp. 21--3.
\newblock
  doi:{\changeurlcolor{black}\href{https://doi.org/10.1007/978-0-387-88136-2_2}{\detokenize{10.1007/978-0-387-88136-2_2}}}.

\bibitem[Metcalfe(1994)]{metcalfe1994book}
Metcalfe, A.
\newblock {\em {Statistics in Engineering : a Practical Approach}}; Chapman \&
  Hall: London New York,  1994.

\bibitem[Winkelmann \em{et~al.}(2020)Winkelmann, Cios, Tokarski, Nolze,
  Hielscher, and Kozie{\l}]{winkelmann2020amat}
Winkelmann, A.; Cios, G.; Tokarski, T.; Nolze, G.; Hielscher, R.; Kozie{\l}, T.
\newblock {EBSD} orientation analysis based on experimental {Kikuchi} reference
  patterns.
\newblock {\em Acta Materialia} {\bf 2020}, {\em 188},~376--385.
\newblock
  doi:{\changeurlcolor{black}\href{https://doi.org/10.1016/j.actamat.2020.01.053}{\detokenize{10.1016/j.actamat.2020.01.053}}}.

\bibitem[rep(2020)]{repo_mn}
Repository with supplementary data for the current manuscript.
\newblock \url{https://github.com/wiai/kikuchi-acentric},  2020.
\newblock Accessed: 2020-11-28,
  doi:{\changeurlcolor{black}\href{https://doi.org/10.5281/zenodo.4118076}{\detokenize{10.5281/zenodo.4118076}}}.

\bibitem[Winkelmann \em{et~al.}(2018)Winkelmann, Nolze, Cios, and
  Tokarski]{winkelmann2018prm}
Winkelmann, A.; Nolze, G.; Cios, G.; Tokarski, T.
\newblock Mapping of local lattice parameter ratios by projective {Kikuchi}
  pattern matching.
\newblock {\em Physical Review Materials} {\bf 2018}, {\em 2},~123803.
\newblock
  doi:{\changeurlcolor{black}\href{https://doi.org/10.1103/PhysRevMaterials.2.123803}{\detokenize{10.1103/PhysRevMaterials.2.123803}}}.

\bibitem[Wright \em{et~al.}(2015)Wright, Nowell, de~Kloe, Camus, and
  Rampton]{wright2015um}
Wright, S.I.; Nowell, M.M.; de~Kloe, R.; Camus, P.; Rampton, T.
\newblock {Electron imaging with an EBSD detector}.
\newblock {\em Ultramicroscopy} {\bf 2015}, {\em 148},~132--145.
\newblock
  doi:{\changeurlcolor{black}\href{https://doi.org/10.1016/j.ultramic.2014.10.002}{\detokenize{10.1016/j.ultramic.2014.10.002}}}.

\bibitem[Brodusch \em{et~al.}(2018)Brodusch, Demers, and
  Gauvin]{brodusch2018joi}
Brodusch, N.; Demers, H.; Gauvin, R.
\newblock {Imaging with a Commercial Electron Backscatter Diffraction ({EBSD})
  Camera in a Scanning Electron Microscope: A Review}.
\newblock {\em Journal of Imaging} {\bf 2018}, {\em 4},~88.
\newblock
  doi:{\changeurlcolor{black}\href{https://doi.org/10.3390/jimaging4070088}{\detokenize{10.3390/jimaging4070088}}}.

\bibitem[Winkelmann(2018)]{xcdskd_bse}
Winkelmann, A.
\newblock {\em {xcdskd: Tools and Methods for Kikuchi Diffraction}},  2018.
\newblock \url{https://xcdskd.readthedocs.io/en/latest/bse_imaging.html},
  accessed 24/03/2021,
  doi:{\changeurlcolor{black}\href{https://doi.org/10.5281/zenodo.3689161}{\detokenize{10.5281/zenodo.3689161}}}.

\bibitem[Esling \em{et~al.}(1980)Esling, Bunge, and Muller]{esling1980jdp}
Esling, C.; Bunge, H.; Muller, J.
\newblock {Description of the texture by distribution functions on the space of
  orthogonal transformations. Implications on the inversion centre.}
\newblock {\em Journal de Physique Lettres} {\bf 1980}, {\em 41},~543--545.
\newblock
  doi:{\changeurlcolor{black}\href{https://doi.org/10.1051/jphyslet:019800041022054300}{\detokenize{10.1051/jphyslet:019800041022054300}}}.

\bibitem[Bunge \em{et~al.}(1980)Bunge, Esling, and Muller]{bunge1980jac}
Bunge, H.J.; Esling, C.; Muller, J.
\newblock The role of the inversion centre in texture analysis.
\newblock {\em Journal of Applied Crystallography} {\bf 1980}, {\em
  13},~544--554.
\newblock
  doi:{\changeurlcolor{black}\href{https://doi.org/10.1107/s0021889880012757}{\detokenize{10.1107/s0021889880012757}}}.

\bibitem[Morikawa \em{et~al.}(2020)Morikawa, Yamasaki, Kanazawa, Yokouchi,
  Tokura, and Arima]{morikawa2020prm}
Morikawa, D.; Yamasaki, Y.; Kanazawa, N.; Yokouchi, T.; Tokura, Y.; Arima, T.
\newblock Determination of crystallographic chirality of {MnSi} thin film grown
  on {Si}(111) substrate.
\newblock {\em Physical Review Materials} {\bf 2020}, {\em 4},~011407.
\newblock
  doi:{\changeurlcolor{black}\href{https://doi.org/10.1103/physrevmaterials.4.014407}{\detokenize{10.1103/physrevmaterials.4.014407}}}.

\end{thebibliography}

%%%%%%%%%%%%%%%%%%%%%%%%%%%%%%%%%%%%%%%%%%
\end{document}